# Kinetics and the crystallographic structure of bismuth during liquefaction and solidification on the insulating substrate

Tjeerd R.J. Bollmann[1,2] and Maciej Jankowski[3]
[1]Saxion University of Applied Sciences, P.O. Box 70.000, 7500KB Enschede, the Netherlands
[2]University of Twente, Inorganic Materials Science, MESA+ Institute for Nanotechnology, P.O. Box 217, NL-7500AE Enschede, The Netherlands
[3]ESRF-The European Synchrotron, 71 Avenue des Martyrs, 38000 Grenoble, France

Here we study the kinetics of liquefaction and solidification of thin bismuth films grown on the insulating substrate by the pulsed laser deposited (PLD) and molecular beam epitaxy (MBE) and investigated by *in situ* electron and X-ray diffraction. By PLD, we can grow films similar to those obtained using MBE, studied by ex-situ AFM, KPFM, XRR, and XRD. The liquefaction-solidification transition is monitored in real-time by RHEED and synchrotron XRD, resulting in a dewetting phenomenon and the formation of spherical droplets which size depends on the initial film thickness. Studying this phase transition in more detail, we find instantaneous liquefaction and solidification, resulting in formation of the nanodots oriented with a (110) crystallographic plane parallel to the substrate. Furthermore, we propose a two-step growth mechanism by analyzing the recorded specular diffraction rods. Overall, we show that the PLD and MBE can be used as a method for the highly controlled growth of Bi nanostructures, including their crystallographic orientation on the substrate.

**Introduction:**

The solid-state dewetting on solid substrates is one of the fundamental processes that understanding is essential for designing high-performance devices based on thin films [1]. Thin films deposited on a substrate are usually in the metastable phase, and their solid-liquid-solid transition, driven by a minimization of the free energy of the system, is irreversible and leads to more energetically stable surface morphologies. However, the final configuration of the surface is disparate from its initial state and typically results in the formation of islands by agglomeration. Although dewetting of thin films in many cases is undesired, identifying new phases and factors driving system kinetics during melting and re-solidification at the nanoscale offers new possibilities utilizing the controlled dewetting. The study of these phenomena, continuously improving over the last decades, has high technological importance [2]. Furthermore, the possibility of gaining control over dewetting and solidification of thin films offers novel approaches in the bottom-up engineering of a new class of nanomaterials [3].

Thin bismuth films, studied intensively in recent years, exhibit a plethora of interesting physical properties, including magnetoresistance, metal-insulator transition as a function of film thickness, diamagnetism, and quantum size effects (QSE). Multiple studies are reporting Bi growth on Si crystal surfaces [4–13], HOPG [14–17], as well as other surfaces [18–26], seeking for deeper understanding and control of growth as the resulting films differ in their morphology, orientations, and strain. Exhibiting exotic magneto-electronic properties, Bi films are appealing candidates for spintronic applications [24–35], especially topologically insulating Bi films, members of a novel and promising materials class, due to their spin-momentum locked surface states [17,36–38]. We recently described the controlled growth of high-quality Bi(110) and Bi(111) films on an insulating substrate [39] as a viable route towards practical applications. The growth of high-quality films, a necessity to not hamper their electronic properties due to bulk conduction [31,40,41], on a well-defined insulating substrate that provides an infinite potential well barrier, allows excluding the electronic interference between the film and substrate states [34]. However, controlled growth of thin metal films with well-defined morphology and crystallographic orientation on semiconductor and oxide surfaces is not a trivial task as it typically exhibits rough 3D growth [42], whereas atomically smooth (2D) films are desired. As we recently described, this limitation can be overcome by using low-temperature growth, and metastable crystal structures can be grown and smoothed up to the melting temperature of the film [39]. For the preservation of electronic properties, knowledge about the thermal stability of thin films is a necessity. Bismuth is an exemplary element, which has attracted particular attention in phase-transition studies thanks to its low melting temperature, supercooling far below its melting point [43–45], and size-dependent melting [43]. However, the film morphology and crystallographic orientation behavior crossing the first-order phase transition between liquid and solid is not beforehand predictable, characterization of the nucleation and growth process during melting and freezing remains technically challenging as it requires in situ experimental methods [46,47] allowing to monitor strain, crystallographic structure, and morphological changes during phase transition. Moreover, in typical ensembles, (collective) phase transition events may be rapidly stringing along, even in parallel, thereby possibly screening the individual process. Therefore, investigation of individual islands/nanoparticles is essential to study the phase transition in detail and can help to identify routes towards the controlled formation of nanostructures, i.e., size-depended melting, coverage-dependent re-crystallization [48], Bi nano-alloys [49], and nanocrystals [44]. Bi-based nanomaterials have been recently reported to have enhanced photo-response activity, revealed in Bi quantum-nanodots [50] or Bi nanosheets [51] used as high-performance UV-VIS photodetectors. This work describes a combination of methods to study growth and the (irreversible) phase transition of Bi films, grown by Pulsed Laser Deposition (PLD), into Bi nanodots. PLD is a physical vapor deposition technique where the typical high supersaturation is used to steer smooth thin film growth and even allows to tune the crystallographic orientation by choosing the appropriately used laser fluence [52] at RT, in contrast to growth by MBE at RT [39].

We demonstrate the growth of high-quality Bi film at RT by PLD, similar to films grown by MBE [39], characterized by *in situ* Reflective High Energy Electron Diffraction (RHEED), *in situ* X-ray Photoelectron Spectroscopy (XPS), *ex situ* X-ray diffraction (XRD) and reflectivity (XRR), Scanning Electron Microscopy (SEM) together with *ex situ* Tapping Mode Atomic Force Microscopy (TM-AFM) and Frequency Modulated Kelvin Probe Force Microscopy (FM-KPFM). By analyzing the obtained information, we can determine the post-growth morphology changes. Obtained results are supplemented by *in situ* and real-time synchrotron surface X-ray diffraction (SXRD) measurements, allowing to determine the crystallographic transition during liquefaction and solidification and describing the mechanism of Bi nanodots nucleation. This paper is organized as follows. Section II describes the used experimental conditions for growth by PLD and MBE and the technical details of the used analysis equipment. Section III describes the obtained results for the grown PLD films compared to the MBE grown films. In section IV, we discuss the model for nucleation and growth, driven by the reduction of surface free energy. Finally, in section IV, we summarize and draw conclusions for the liquefaction and solidification of thin Bi films.

**Experimental:**

For the laboratory experiments described here, we used single-crystal α-Al$_2$O$_3$(0001) substrates of 5x5 mm$^2$ having a miscut <0.2□. Before annealing for 12 h in a tube furnace at 1323 K using an O$_2$ flow of 150 l/h, the samples were ultrasonically degreased in acetone and ethanol. The samples were initially inspected by tapping mode atomic force microscopy (TM-AFM) for their step height and terrace width and by XPS to verify the surface cleanliness where only minor traces of C and Ca were found, as described in the literature [39].

After insertion into the UHV setup, thin Bi films were grown by pulsed laser deposition (PLD) in a Twente Solid State Technology (TSST) system with a base pressure of 1x10$^{-8}$ mbar. The growth was monitored by in situ reflective high-energy electron diffraction (RHEED), operated at pressures up to 0.3 mbar, enabled due to differential pumping [52]. A Bi target of 99.999% purity was ablated with a typical energy density of 1.4 J/cm$^2$ and a repetition rate of 0.25 Hz while the substrate was held at RT in the vacuum environment. For all depositions, the target-substrate distance was fixed at 50 mm. After deposition, the samples were heated at 10 K/min towards 600 K, where heating was stopped as soon as the RHEED diffraction pattern vanished to prevent desorption of the film. After deposition, the thin films were slowly cooled down to RT at a rate of 10 K/min. Topographic and local contact potential features of the samples were measured at room temperature by *ex situ* tapping mode (TM) and FM-KPFM mode, respectively, on a Bruker Dimension ICON microscope using Pt/Ir coated probes (Nanosensors PPP-EFM-10). XPS measurements were acquired *in situ* on an Omicron Nanotechnology GmbH setup, equipped with a monochromatic Al K-alpha X-ray source (1486.6 eV) and an EA 125 electron analyzer. All spectra were acquired in the constant analyzer energy (CAE) mode. Crystal structure characterization was done on an X'PertPowder X-ray diffractometer from PANalytical. For the surface X-ray diffraction (SXRD) experiments described here, we used hat-shaped α-Al$_2$O$_3$(0001) single crystals with a miscut of <0.2 □. Before insertion into the experimental UHV setup, the samples were cleaned and analyzed in the same manner as described above. After insertion into the UHV system of the surface diffraction beamline ID03/ESRF (Grenoble, France) [53] with a base pressure below 1x10$^{-10}$ mbar, the sample was cleaned by mild 700 eV Ar+ sputtering at p(Ar)

= $3\times10^{-6}$ mbar and subsequently annealed to 1200 K in $O_2$ pressure of $1\times10^{-6}$ mbar. The sample chemical composition was monitored using Auger electron spectroscopy (AES), see elsewhere [39]. Bi was deposited at a typical deposition rate of 1.3 Å per minute from a Mo crucible mounted inside an electron-beam evaporator (Omicron EFM-3). According to the bulk phase diagram, Bi and sapphire are immiscible in bulk [54]. The SXRD experiments were performed using a monochromatic synchrotron X-ray beam at 24 eV and a four-chip MAXIPIX detector [55]. For data integration and reconstruction of the reciprocal space maps from the 2D detector frames, we used the BINoculars software package [56]. Some of the 2D detector frames are displayed by a colormap that is developed for the screen display of intensity images to be monotonically increasing in terms of its perceived brightness for clarity [57]. All reciprocal space positions are given in (h,k,l) measured in reciprocal lattice units (r.l.u.) of the substrate hexagonal surface unit cell. Bragg peaks of the thin Bi films are labeled by their conventional rhombohedral Miller indices [27]. X-ray reflectivity (XRR) curves have been fitted using the GenX software package [58].

## Results

### *PLD deposition*

In order to study thin Bi films with low surface roughness, we use PLD, a well-studied technique to aid in the synthesis of multicomponent and complex thin films. Compared to conventional physical vapor deposition (PVD) techniques, PLD provides additional parameters, thereby giving more flexibility to the (control of the) growth process of thin films. Besides this, PLD is known for a high supersaturation of target material during growth, resulting in a high nucleation density and inducing the growth of (ultra)smooth thin films [59]. However, although much work is done on the PLD growth of (complex) oxides [59], only a few studies are done on the initial stages of the growth of metallic thin films [60–65].

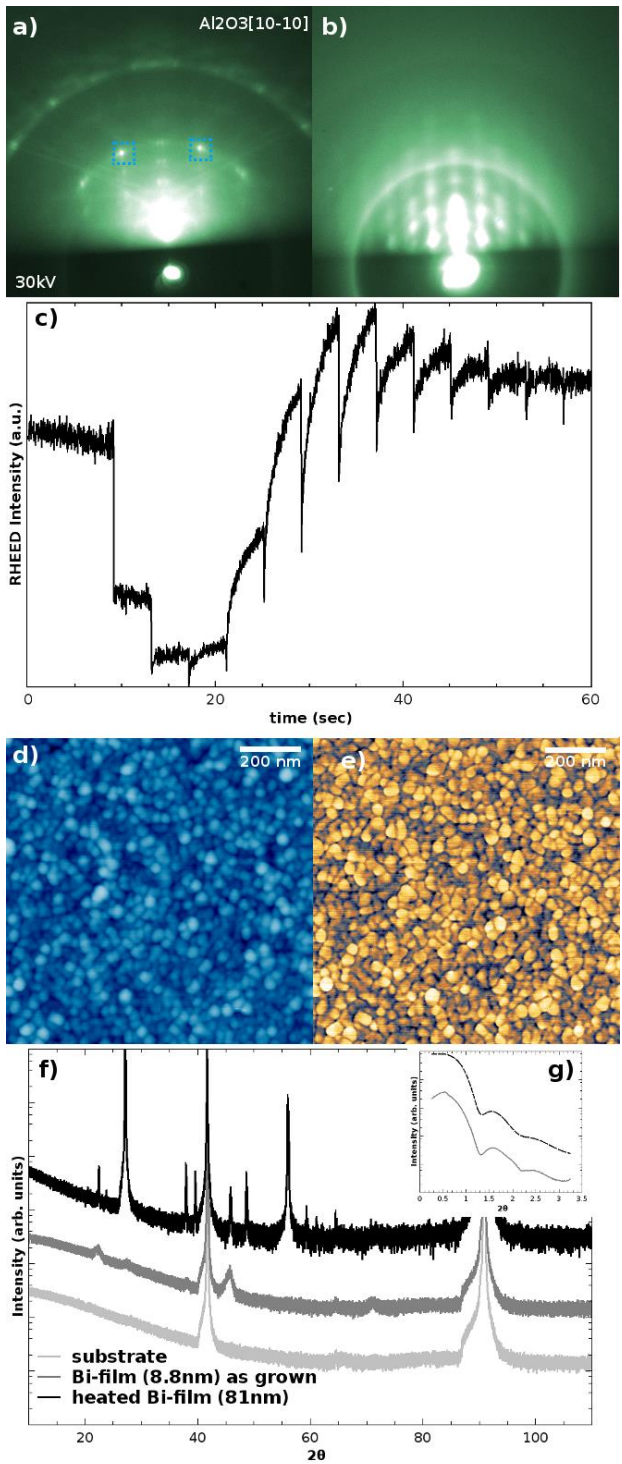

*Figure 1: (color online) (a) RHEED pattern of the sapphire(0001) substrate. Dashed blue squares mark spots due to reflective diffraction of the sapphire substrate for a beam incidence along the [1010] direction of the sapphire substrate. (b) Transmission RHEED pattern of the as-deposited*

*16 nm Bi film. (c) RHEED oscillations during deposition started at t=9 seconds. (d) TM-AFM image of an 8.8 nm grown Bi film revealing an average grain size of 12 nm. The determined surface RMS is 0.85 nm. (e) the corresponding TM-AFM phase-lag image. (f) X-ray diffraction curve for the pristine $Al_2O_3$(0001) substrate and the as-grown 8.8 nm Bi film, revealing its Bi(111) crystal orientation. The diffraction curve shows a predominantly Bi(110) crystal orientation upon melting and cooling. (g) The X-ray reflectivity curve for the as-grown 8.8 nm thick Bi(111) film (full curve). The dashed curve shifted for clarity has been obtained by fitting, resulting in a film thickness of 8.8 nm and surface roughness of 1.2 nm.*

For in-situ characterization during growth, we used RHEED, which reveals a clean sapphire(0001) pattern before deposition, as shown in Fig. 1(a) [66], from which we can calibrate the measured distance between the dashed blue squares as twice the interatomic distance of 2.75Å on the (0001) surface of sapphire. Upon deposition of Bi at RT, a continuous decay in the intensity of the sapphire spots is observed. For thicker films, beyond 4 nm, clear transmission RHEED patterns [67] of Bi appear; see, e.g., the RHEED pattern for a 16 nm thick film in Fig.1(b). Note that the bright ring around the direct beam is a halo feature due to the RHEED arrangement. The transmission pattern was checked by tilting the sample, thereby varying the angle of incidence with respect to the surface. By this, substrate features should reveal position changes as they are very sensitive with tilt, in contrast to transmission features that keep their position. Upon deposition of Bi at RT, we find intensity relaxations in the signal of the specular spot recorded during deposition, as plotted in Fig. 1(c). From this, we find an exponential rise for the recovery of the RHEED signal after each pulse, where we can fit Eq. 1 of Ref. [68] to find a characteristic diffusion time τ ranging from 2.3 seconds down to 0.7 seconds for increased coverage of 4 to 12 ML.

Investigating the morphology of the as-grown 8.8 nm thick Bi film by TM-AFM, see Fig. 1(d) and (e), reveals a grainy film having an average grain size of about 12 nm. The observed surface roughness ($\sigma_{rms}$) is only 0.85 nm. In Fig. 1(f), we plot the X-ray diffraction curve for the pristine sapphire substrate and the as-grown 8.8 nm thick Bi film. From the broadening of the Bi(111) and Bi(222) diffraction peaks at respectively θ=21.6º and θ=45.5º, we find by applying the Scherrer equation (K=0.89) an average crystallite size of respectively 15 nm and 11 nm in good agreement with our real-space observations in Fig. 1(d) and (e). From the X-ray reflectivity scan, the exact film thickness is obtained, from which we determine the deposition rate per pulse to be 0.088 nm/pulse. A fit to this reflectivity scan gives a surface roughness of 1.2 nm, in accordance with the topographic TM-AFM observations in Fig. 1(d).

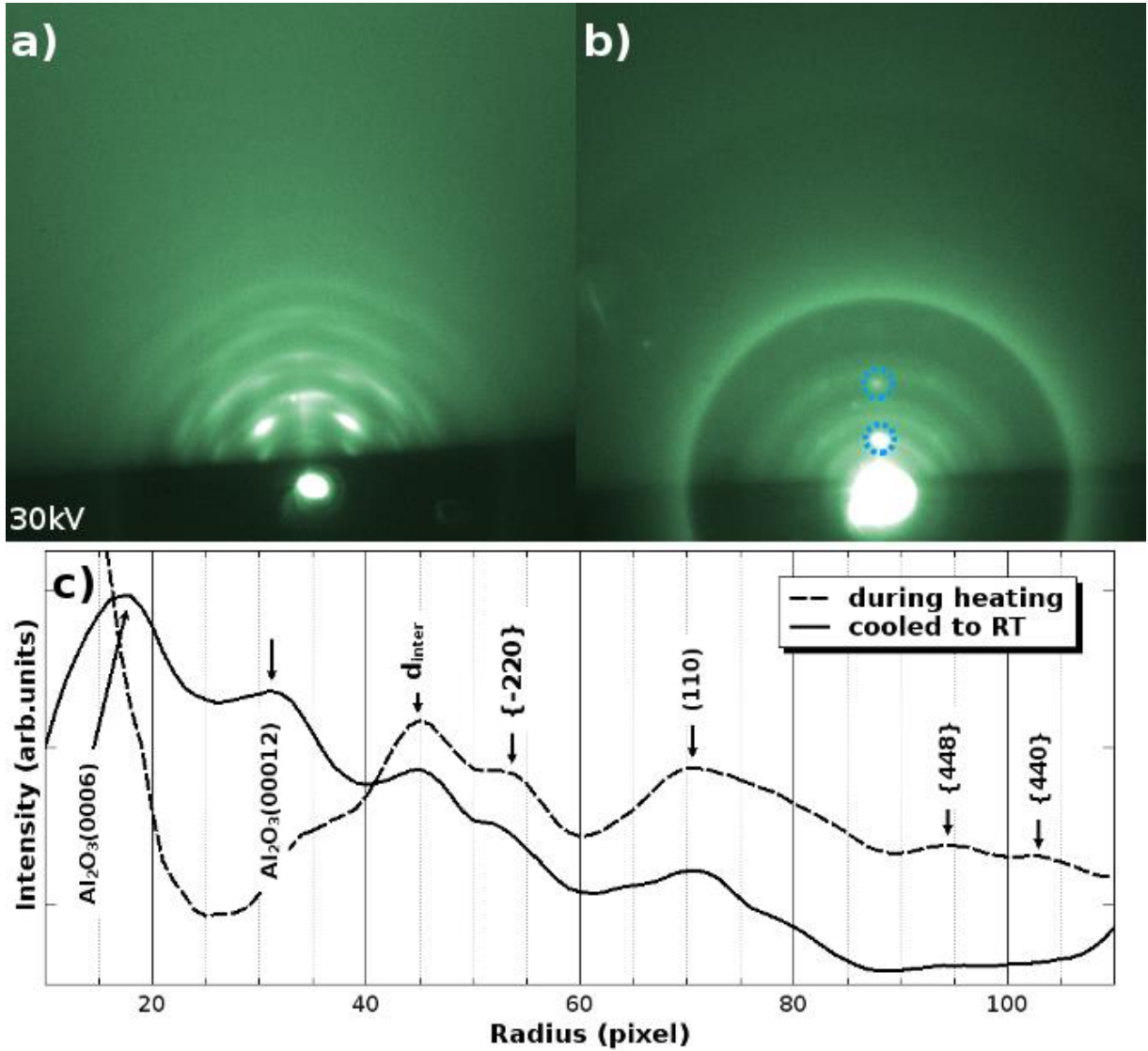

*Figure 2: (color online) (a) RHEED pattern of a heated Bi film where angular dispersion in the textured film orientation increases. (b) Debye-Scherrer rings appear in the RHEED pattern of the cooled Bi film. Dashed blue circles mark spots representing the reflection pattern spots from the free substrate areas. (c) The integral of the radial intensity distribution of Bi nanocrystals after temperature treatment. The labeled arrows indicate the expected positions and intensities of the Debye-Scherrer rings calculated from powder diffraction intensities.*

Upon post-annealing this film at a 10 K/min rate, the RHEED spots slightly show more diffuse spots moving towards a ring-like pattern, see Fig. 2(a). This indicates misalignment and randomness in the crystallites as the angular dispersion in the textured films increases. Upon liquefaction of the Bi film, the RHEED pattern vanishes, and the heating is stopped. Upon cooling, Debye-Scherrer rings appear, indicative of randomly oriented nanoparticles. The rings get more and more intense upon cooling towards RT. The rings exhibit homogeneous intensity along their periphery (see also Supplemental Material), indicative of the isotropic orientation of the nanoparticles on the sapphire substrate. The resulting RHEED pattern at RT, see Fig. 2(b), comprises an RHEED transmission pattern resulting from the particles, together with reflection pattern spots from the substrate areas free of nanoparticles, marked by dashed circles. As said, by tilting the sample with respect to the incident electron beam, the spots corresponding to the substrate are sensitive to sample tilting, whereas the Debye-Scherrer ring positions are insensitive to it. In order to calculate the preferred facets of the nanoparticles, the preferred Debye-Scherrer ring radius is calculated by integrating the radial intensity distribution along the periphery of circles around the RHEED central spots for increasing radius, see also Supplemental Material. The relative ring positions are calibrated by the interatomic distance plane visible for the clean sapphire(0001) substrate in Fig. 1(a). Fig. 2(c) shows the integral of the radial intensity distribution of Fig. 2(a), during heating, and Fig. 2(b), cooled to RT. Upon slight annealing, the substrate RHEED spots of sapphire can only slightly be distinguished. The RHEED spots marked by the dashed circle in Fig. 2(b) clearly contribute to the radial intensity distribution in Fig. 2(c), resulting in pronounced peaks at a radius of 17 and 32 pixels. The corresponding peaks can be related to the Bi(110) Bragg peak at a radius of 69 pixels. The interatomic distance of (001) Bi planes, 3.28 Å apart, corresponds to the peak at 45 and 82 pixels radius. In order to determine the realspace morphology changes corresponding to change in the reciprocal RHEED pattern, as seen in Fig. 2(a-b) and discussed above, we image the resulting morphology by ex-situ TM-AFM as shown in Fig. 3 for decreasing film thicknesses. Note that the resulting (average) film thickness is determined using calculations of the total volume from representative AFM images, as during heating, some of the material might desorb from the substrate resulting in a lowered (average) film thickness. Remarkably, we find spherical cap-shaped grains whose (average) size depends on the initially grown film thickness. This method might be a viable route towards the growth of Bi nanodots with controlled nanodot size. Note that both the as-grown film and the solidified spherical cap-shaped grains reveal no chemical bonding to the oxygen lattice of the substrate surface as verified by XPS in the Supplemental Material Fig. 15. The Bi $4f_{5/2}$ and $4f_{7/2}$ peaks show no shifting, indicative of $Bi_2O_3$ bonds. Although the net bonding enthalpy to form the Bi-O bond is only 137 kJ/mol [69], the bonding enthalpy of the Al-O bond is 511 kJ/mol, corroborating the chemical stability of the substrate.

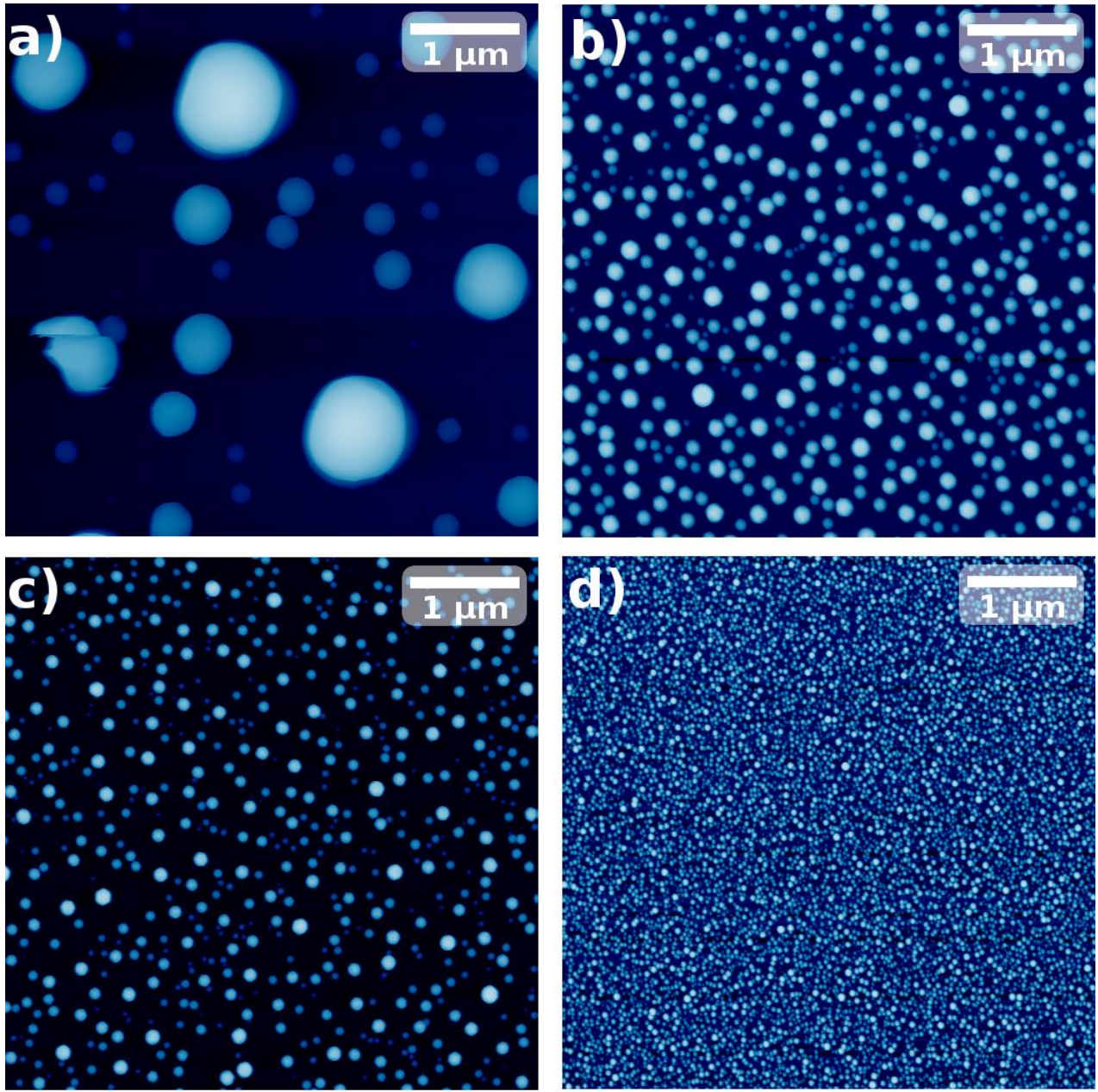

*Figure 3: (color online) TM-AFM topographic images of cooled Bi films having different (average) film thicknesses of 81.2 nm (a), 15.2 nm (b), 7.8 nm (c), and 3.8 nm (d).*

To get a deeper understanding of the melting process proceeding the liquefied Bi film, we heated a PLD grown Bi film just towards melting but abruptly stopped the heating power upon the first changes in RHEED intensity and pattern. The resulting film was then analyzed by AFM measurements, as shown in Fig. 4(d). In this morphology, we identify both Bi spherical caps, as already discussed in Fig. 3(a-d), and hut-shaped Bi crystals. As the Bi films appear to be uninfluenced by the registry of the substrate, this Wulff crystal [70] shapes show random tilt. Line profiles illustrate this tilt, the flatness, and the hut shape along with the crystal, as shown in Fig. 4(e). A detailed analysis of one such crystal is shown in Fig. 4(a-c), where we investigated the crystallite corresponding to the dashed box in Fig. 4(d). Dashed lines mark the shape of the crystal as in the amplitude image, the shape is easily identified. Measuring the local contact potential difference (LCPD) by FM-KPFM, one can quickly identify the different workfunctions of the Bi facets resulting from the different stacking. As the crystals all show a random tilt with respect to the substrate, these facets can not easily be uniquely identified.

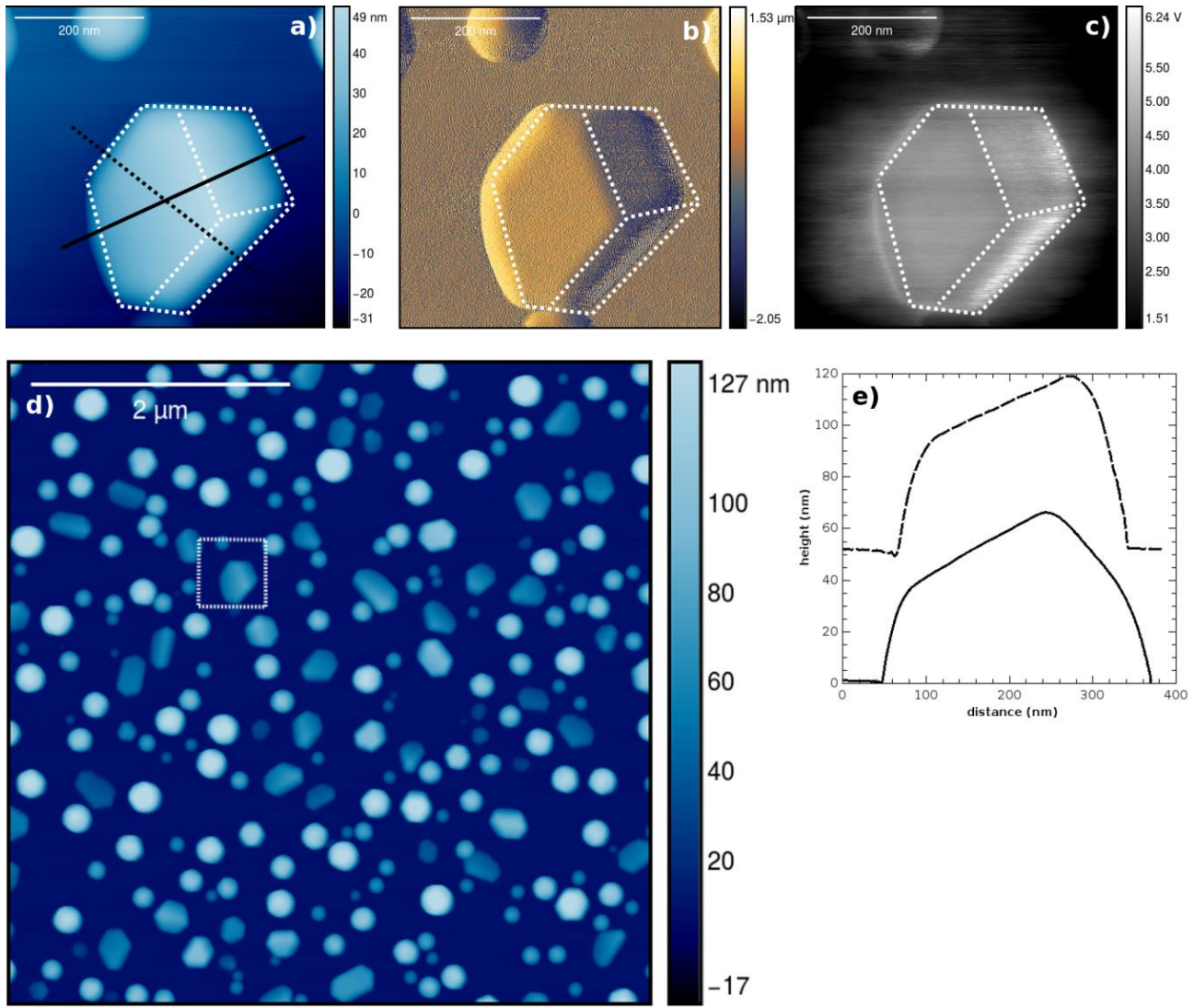

*Figure 4: (color online) TM-AFM images of a single hut-shaped crystal in topography (a), amplitude (b), and LCPD (c). (d) TM-AFM overview image (5x5µm²) showing a quenched melting process where dewetting resulted in many hut-shaped Wulff crystals as well as many hemispherical Bi grains. The area marked by the dashed square corresponds to Fig. 4(a). (e) The line profiles are taken along the corresponding full and dashed line in Fig. 4(a).*

## *MBE deposition and synchrotron studies*

In order to get a deeper understanding of the transition involved in solidification, we need to study this phase transition in-situ determining crystallographic orientations and morphology changes. As typical phase transition events might be, even in parallel, rapidly stringing along, individual processes are hard to study although essential to get a deeper understanding. As this transition is hardly accessible by standard laboratory equipment available and as we need to be very surface sensitive, we extend our observations by Surface X-ray diffraction (SXRD) using synchrotron-based X-ray radiation.

Our previous research described the growth and crystallographic phase evolution with the temperature of thin Bi films on the sapphire substrate using synchrotron SXRD [39]. In short, to obtain high-quality Bi films, deposition at 40 K is necessary in order to limit Bi kinetics and obtain uniformly thick films with the random azimuthal orientation parallel to the $Al_2O_3$(0001) substrate. First, those films are oriented, having their (110) crystallographic plane parallel to the substrate. Subsequent annealing up to 300 K causes thin-film smoothing, and at 400 K, the crystallographic phase transition to (111) orientation is observed. Further heating, up to 500 K, results in ultra-smooth Bi films with reduced roughness [39].

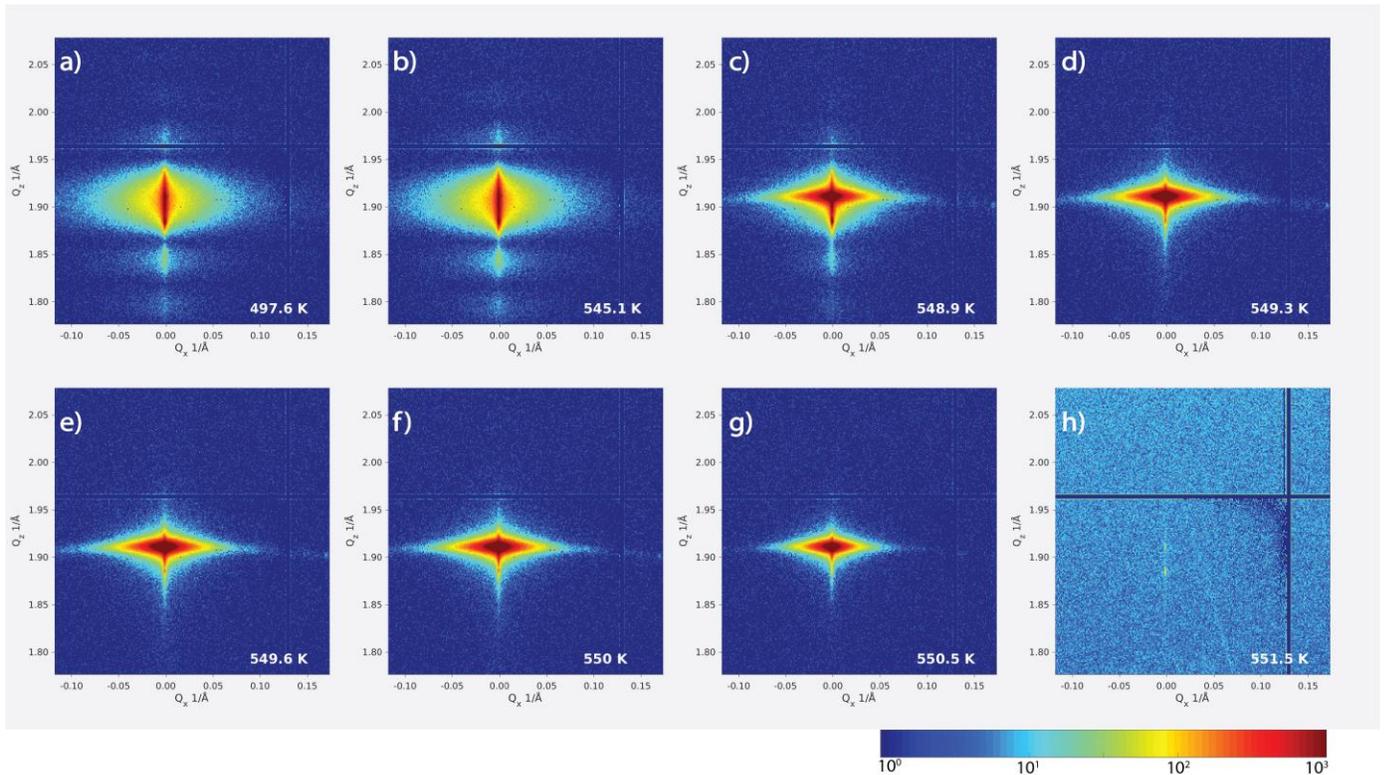

*Figure 5: Detector snapshots recorded during melting of a 14 nm thick Bi film. The position of the detector was set on the Bi(111) Bragg peak. The intensity of the recorded signal is presented on a*

*logarithmic color scale.*

To monitor the evolution of the thin film morphology and the expected solid-liquid phase transition at the melting point of Bi, we grow a 14 nm thick film on Al$_2$O$_3$(0001) [39]. Subsequent heating and cooling of this film provide deep insight into the kinetics involved during liquefaction and solidification. The typical intensity profile of (00L) crystal truncation rod (CTR) recorded from a Bi film prepared as such is presented in Fig. 7(c). The Bi(111) Bragg peak at L=3.3 is surrounded by pronounced fringes covering the whole range of L and evidencing the high quality of the film having an average roughness of only a few Å [39] and oriented with the (111) plane parallel to the substrate.

In order to follow the melting and liquefaction of the thin film, we positioned the detector at the Bi(110) Bragg peak, where the plane of the detector cuts through the center of the peak and neighboring Laue fringes, as it is shown in Fig. 5(a) and Supplementary Movie M1. Gradual increasing the sample temperature, with a slow rate of 1 K/s, does not cause changes in the observed pattern and the peak intensity, plotted in Fig. 5(a), until 548 K at which concomitantly the Bragg peak suddenly narrows, increases in the intensity, and Laue fringes suddenly disappear (Fig. 5(c)). We attribute this rapid increase of the peak intensity to the de-wetting of the film, which transforms to large crystallites of Bi in a very narrow temperature window (548-550 K), as can be seen in Fig. 5(c)-(g) just before liquefaction as is detected in Fig. 5(h) where the Bragg peak has disappeared. The increased ordering of the Bi structures is revealed as the peak intensity of the Bragg peak is increasing while its width in the out-of-plane ($k_z$) direction is decreasing. At 550 K, the Bragg peak suddenly disappears, followed by an increase of the background level arising from the diffuse scattering of liquid Bi (Fig. 5(h)), as the weak reflection of the substrate specular rod is present. The melting temperature of the Bi films is at 550 K, slightly above the bulk melting temperature of Bi at 545 K that is reported in the literature by Takagi et al. [43].

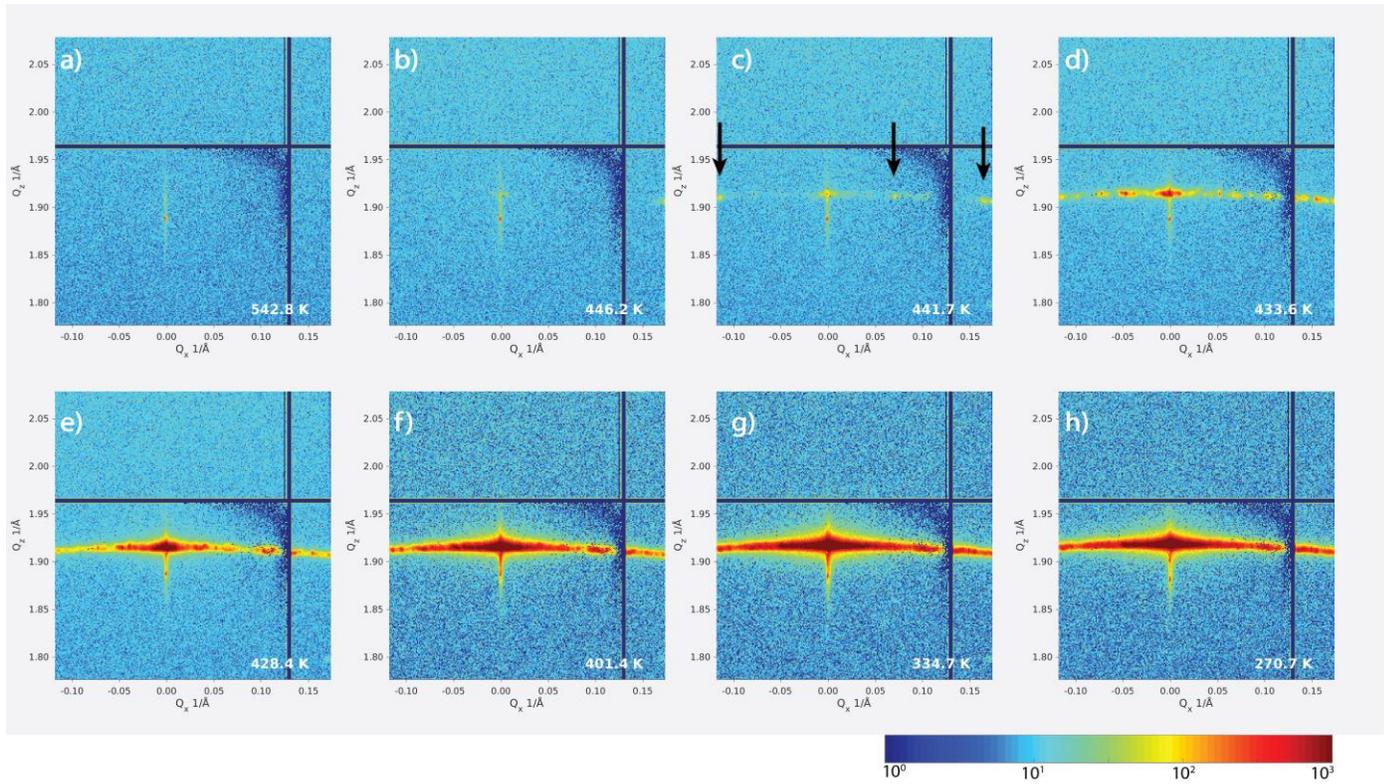

*Figure 6: Detector snapshots recorded during solidification of the liquid Bi film. The detector was positioned at the Bi(110) Bragg peak. The arrows in (c) indicate the reflexes originating from the nucleating Bi crystallites. The intensity of the recorded signal is presented on a logarithmic color scale.*

As we already determined the crystallographic structure for the spherical caps observed upon cooling of the liquefied Bi film to be Bi(110), we, therefore, trace the shape and intensity at the Bi(110) Bragg peak position, see Fig. 6(a) and Supplementary Movie M2, while cooling the sample towards RT at a slow rate of 0.05 K/s, the intensity of the peak as a function of temperature is plotted in Fig. 7(b). At the temperature of 440 K, we see an onset of the weak intensity, then rising up with a temperature up to 370 K, where it saturates. A quick check by an (00L) CTR scan confirms the conversion of the Bi(111) crystal structure before liquefaction towards Bi(110) crystal structures upon solidification, see Fig. 7(c) and (d). The peak at L =3 in Fig. 7(c) is the forbidden reflection $Al_2O_3$(0003) and arises from the presence of defects in the bulk of the substrate crystal.

A closer look at the Bi(111) peak shape reveals dramatic change if compared with the Bi(110) Bragg peak before the melting transition. The first changes in the recorded images are observed in Fig. 6(c) at around 440 K, which is 110 K lower than the melting point of the Bi film, visible as weak reflexes appearing around the Bi(111) peak position, along the (110) diffraction ring. Those reflexes originate from large crystallites (a few hundred nanometers in size) solidifying from the supercooled liquid phase of Bi (see Supplementary Movie M2). Their position deviates from the Bi(110) peak position due to a slight tilting of (110) planes of the nucleated crystallites with respect to the substrate. Further temperature decreases lead to the spontaneous formation of more crystallites, Fig. 6(d)-(g). At room temperature, most of the measured intensity is concentrated in the Bi(110) peak lying on the weak Debye-Scherrer ring which intensity gradually decreases further away from the peak, see Fig. 6(h).

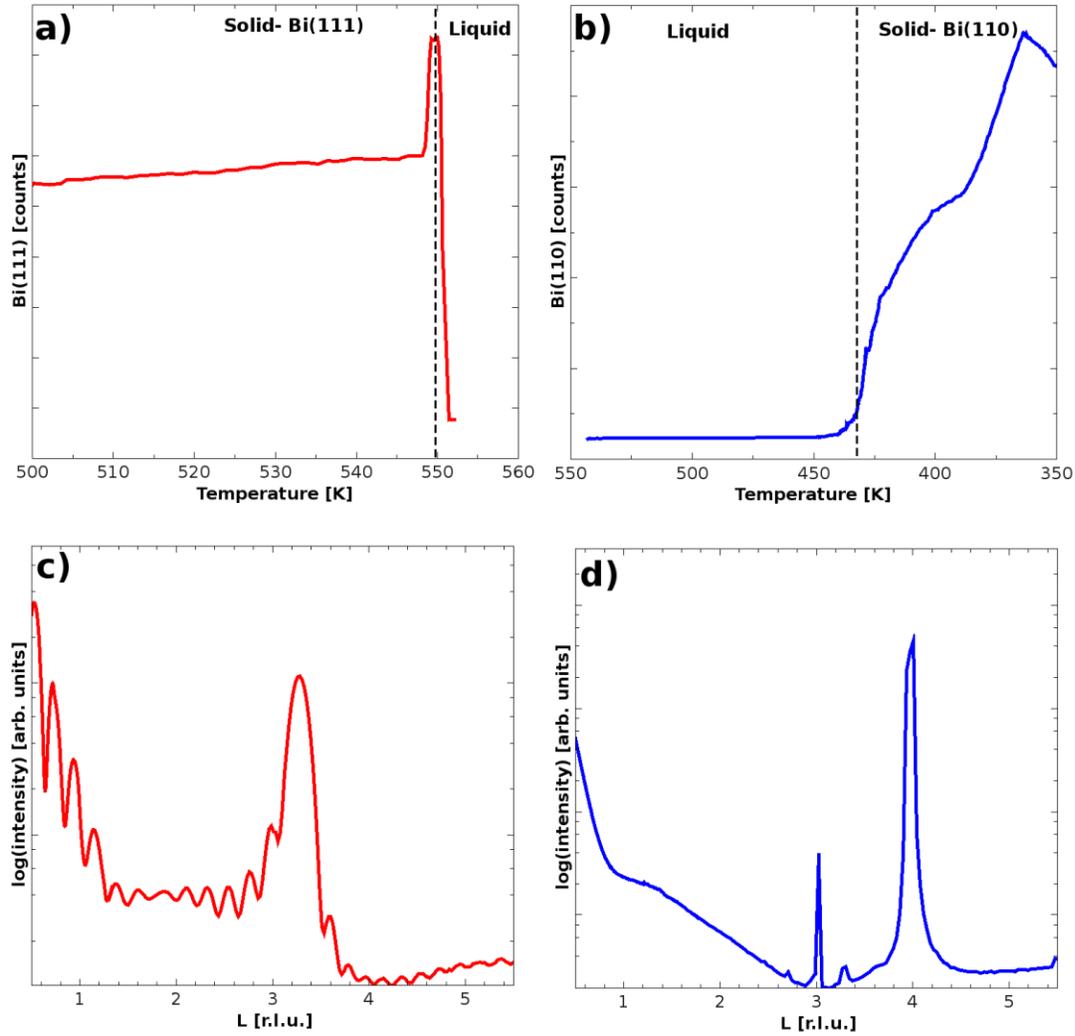

*Figure 7: (color online) TThe X-ray intensity evolution as a function of temperature measured at (a) L=3.3, corresponding to Bi(111) films, and at (b) L=4, corresponding to Bi(110) films. (c) X-ray reflectivity curve of the Bi(111) film before liquification, revealing multiple Laue oscillations ans Kiessig fringes. (d) X-ray reflectivity curve for the Bi(110) crystal structure measured upon solidification of the liquified Bi films upon cooling.*

In order to analyze the growth kinetics, we study the individual frames recorded by the MAXIPIX 2D detector [55] recorded during solidification. In Fig. 8, we display some frames of an individual solidification event. For increasing time, we see the development of the Debye-Scherrer ring onto the 2D detector, see also Fig. 8(d), made up of individual spots appearing over time, having an offset in $k_{//}$. See also Supplemental Material for a full movie showing the analysis of several of these spots, also discussed below.

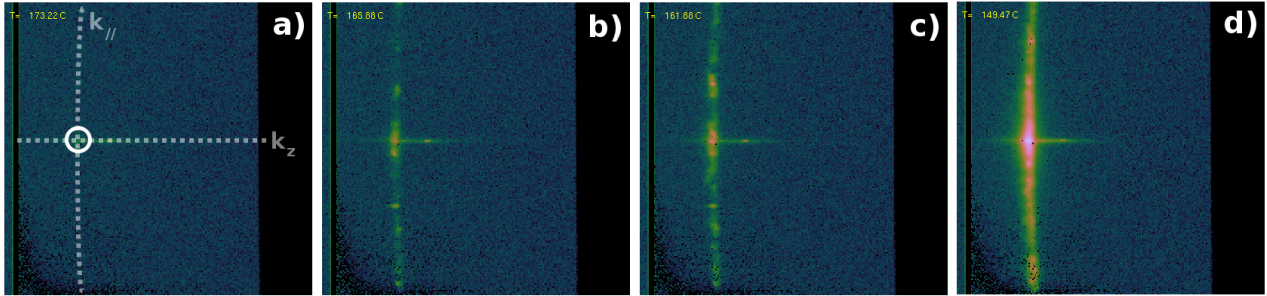

*Figure 8: (color online) Frames of the MAXIPIX 2D detector [50] of 512x512 pixels measured during the recrystallization of Bi at (a) t=10 sec, (b) t=70 sec, (c) t=130 sec, (d) t=290 sec. Horizontally $k_z$ and vertically $k_{//}$ is projected.*

For a more detailed analysis, we measure the intensity of the central spot, see Fig. 8(a) circular marker, as well as several individual spots that appear during recrystallization, as well as their FWHM in the direction of $k_z$ and $k_{//}$ where we assume a Gaussian shape for the intensity. In Fig. 9, we summarize the obtained data from a single solidification event where we plot the FWHM for both $k_z$ and $k_{//}$ as a function of increasing frame number (time) in Fig. 9(a). We see a different behavior for $k_z$, and $k_{//}$ as the FWHM for $k_{//}$ shows an almost discrete onset within several frames up to its maximum of ~4 pixels. However, the FWHM of $k_z$ increases during about 40 frames up to its maximum of a similar value. The greyed box in Fig. 9(a-c) relaxation of the sample holder postion due to the temperature change, influencing the measured intensity.

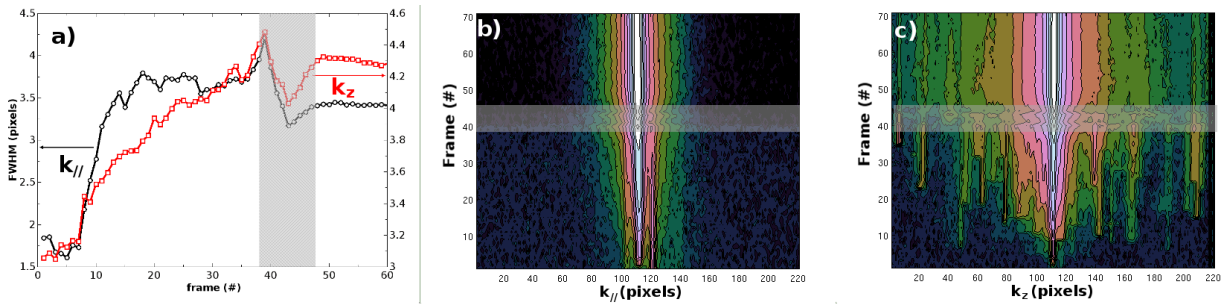

*Figure 9: (color online) (a) The FWHM of the central spot depicted by the circular mark in Fig. 8(a) for in-plane $k_{//}$ (black) and out-of-plane $k_z$ (red) derived from line-profiles through subsequent frames of the MAXIPIX 2D detector. (b) Stacked line profiles for subsequent 2D detector frames*

*through $k_{//}$, see the dashed line labeled $k_{//}$ in Fig. 8(a). (c) Stacked line-profiles for subsequent 2D detector frames through $k_z$; see the dashed line labeled $k_z$ in Fig. 8(a).*

From the literature, it is well known that the FWHM is inversely related to the (average) grain size by the Scherrer equation [71]. The Scherrer equation:

$$FWHM = \frac{K\lambda}{\beta cos(\theta)} \quad (1)$$

relates the FWHM to the average crystal size $\beta$, the grazing incident angle $\theta$, the used X-ray wavelength $\lambda$, and the Scherrer constant $K$. The Scherrer constant $K$ is commonly assumed to be a constant between 0.8 and 1.1. The interpretation of this equation, assuming the Scherrer constant $K$ to be a constant, to our obtained data would counterintuitively suggest a decreasing (average) grain size during solidification, both in-plane ($k_{//}$) as well as out-of-plane ($k_z$) as the FWHM in both, see Fig. 9(a), is increasing. However, the Scherrer constant in this experiment can not be assumed to be constant, as the Scherrer constant [71] is dependend on the dispersion of crystallite sizes [72-73].

To clarify this, we plotted in Fig. S4 how the value of the Scherrer constant ($K$) can be considered to depend on the normalized grain size distribution ($\beta/\sigma$), see also Fig. S4(a). To explain this behavior, we plotted the normalized grain size distribution in Fig. S4(b). During growth, the average grain size ($\beta$) increases. When the (normalized) grain size distribution ($\sigma$) has a constant width, the ratio of $\beta/\sigma$ will increase as illustrated in Fig. S4(b). An ensemble of grains having constant size distribution width ($\sigma$) while growing in average crystallite size ($\beta$) will therefore increase its Scherrer constant ($K$) as illustrated by the grey, black, green, and blue curve in Fig. S4(b) and the corresponding diamond markers in Fig. S4(a). In the case that the grain size distribution width gets narrower during growth, the Scherrer constant will also be increased, as illustrated by the red curve in Fig. S4(b) and corresponding red diamond marker in Fig. S4(a).

Assuming the Scherrer constant ($K$) not to be fixed, we can explain the increasing FWHM both in- and out-of-plane as shown in Fig. 9(a). These two scenarios, (1) the growth of the average crystallite size (*β*) while having a constant size distribution (*σ*) as well as (2) of a size distribution that is narrowing, sketch the increase of the FWHM measured during solidification events we observed. For our measurements in Fig. 9(a), the grain size in-plane is rapidly fixed, indicated by the almost discrete onset of $k_{//}$ as seen in Fig. 8(a). While the growth in-plane is rapidly fixed, the out-of-plane growth continues, increasing *K* by the two scenarios/mechanisms discussed above, resulting in an increasing FWHM, see also Eq. 1, as illustrated in Fig. 9(a).

## Discussion

As seen from the experimental data obtained slightly before a fully melted Bi film, see Fig. 4 we find Wulff shaped crystals and hemispherical Bi grains, dewetting the substrate. Wetting can be described by Young's equation [74]:

$\gamma_{SG} = \gamma_{SL} + \gamma_{LG} cos(\theta_C)$ (2)

where the interplay between the surface free energy of the substrate ($\gamma_{SG}$), the Bi film ($\gamma_{LG}$) and the interface between substrate and film ($\gamma_{SL}$) determines the contact angle ($\theta_C$) for dewetting. As the surface, free energy of Bi is described in the literature as about 0.4 J/m$^2$ (or 2.5 eV/nm$^2$) [75] and the surface free energy of the sapphire substrate about 1.3 J/m$^2$ [76], as an educated guess one would expect the film to wet the substrate, not knowing the role of the interface of Bi and substrate. However, the solidified Bi spherical caps indicate dewetting. When determining the contact angle in detail from the AFM images measured, we find contact angles between 77º<$\theta_C$<86º. Note that by scanning probe microscopy, contact angles beyond 90º cannot be probed due to the tip convolution effect. We therefore verified this by SEM side-views, revealing the contact angles to be corresponding to the measured AFM image, <90º. A representative example can be found in Fig. S3.

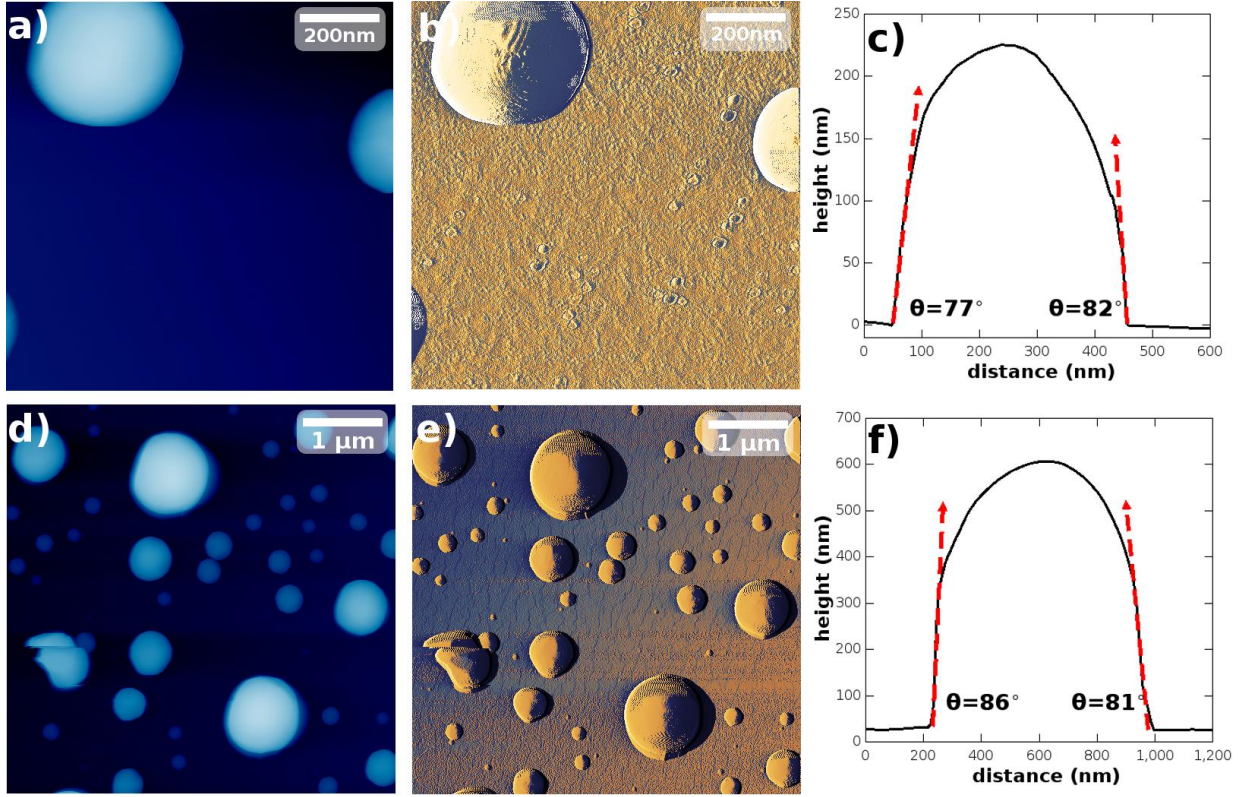

Figure 10: (color online) TM-AFM images of a 14.8 nm, respectively 81.2 nm thick solidified Bi film, topography (a)&(d) and amplitude (b)&(e). (c) & (f) The corresponding line profiles along the center of the spherical caps and the measured contact angles.

Having contact angles close to 90° gives $cos\theta_C \simeq 0$, reducing Eq. 2 to:

$$\gamma_{SG} = \gamma_{SL} \quad (3)$$

Assuming the surface free energy of the Bi/substrate interface and the sapphire substrate to be equal, we can calculate the surface free energy gain as the film dewets the surface. In Fig. 11, we compare the grainy as-grown PLD film revealing the Bi(111) crystal orientation, Fig. 11(a), to the heated and solidified Bi(110) spherical caps resulting from dewetting, Fig. 11(b). For the as-grown film, the individual grains can be modeled as cylinders stacked on the substrate, see inset in Fig. 11(a), as the grain sizes are limited by neighboring grains, where the average grain size is typically determined by the randomly distributed neighboring grains at a radius depending on the PLD supersaturation during growth. The volume of the cylinder is calculated by [77]:

$$V_{cylinder} = \pi R^2 h \quad (4)$$

while the surface area of the top of the cylinder, the exposed area of the grain, is calculated by:

$$S_{cylinder} = \pi R^2 \quad (5)$$

where $R$ is the radius and $h$ the height.

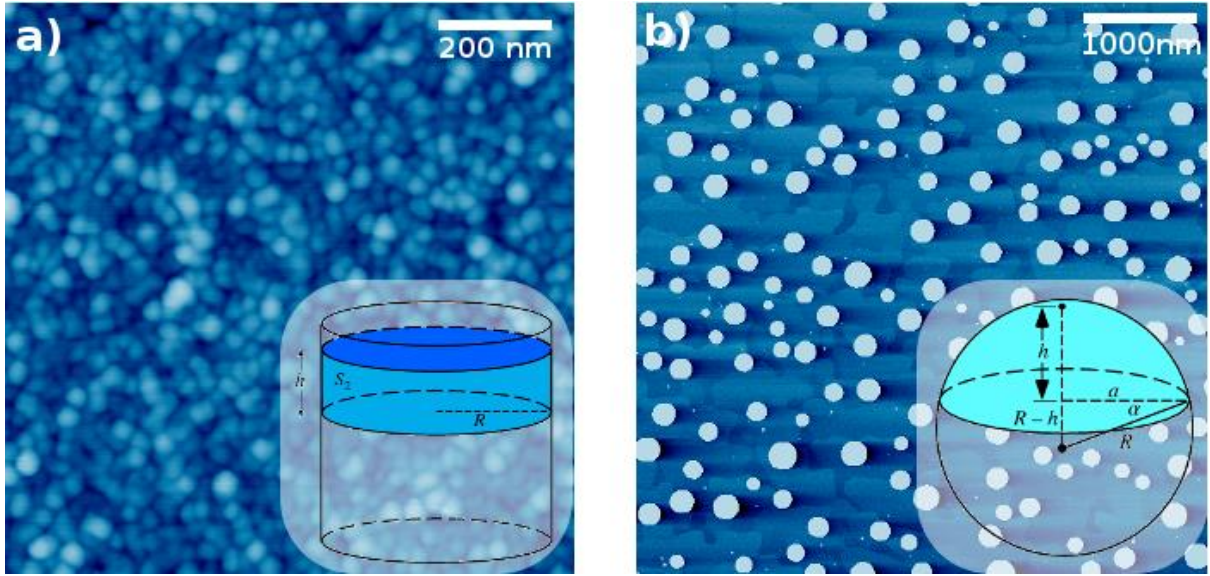

Figure 11: (color online) (a) TM-AFM image of an as-grown Bi(111) PLD film. Inset shows the cylinder geometry used to model the individual grain. (b) TM-AFM image of a solidified Bi(110) (PLD) film. Inset shows the cut sphere geometry used to model the individual Bi spherical caps.

The solidified film, showing the spherical caps with contact angles close to 90º as shown in Fig. 10, can be modeled by a spherical cap as depicted in the inset of Fig. 11(b). Notice the substrate steps visible in Fig. 11(b) by enhancing the image contrast. The volume of the spherical cap is calculated by [69]:

$$V_{cap} = \frac{\pi h}{6}(3a^2 + h^2) \quad (6)$$

and the exposed surface area of the spherical cap by:

$$S_{cap} = 2\pi R h = \pi(a^2 + h^2) \quad (7)$$

where $R$ is the radius of the sphere, $h$ the height of the spherical cap, and $a$ the base radius of the spherical cap.

Now, for the Bi spherical caps in Fig. 11(b), we find by a grain analysis by Gwyddion [78] an average spherical cap height of $<h>$=72 nm and an average base radius of $<a>$=48 nm. From this, the total volume of material can be calculated as $V_{cap}$=456.000 nm³. As the as-grown PLD film has been deposited under the same conditions, we assume the total volume of the deposited material to be the same. As we know from the X-ray reflectivity curve in Fig. 1(g), the average film thickness <h>=8.8 nm, we can calculate by setting $V_{cap}=V_{cylinder}$ the corresponding average grain radius as <R>=128 nm. Now that we have the average dimensions of the grain modeled as a cylinder (*R* and *h*) and the spherical cap (*a* and *h*), we can calculate the reduction of the Bi surface exposed by dewetting. As we concluded previously in Eq. 3, the interface free energy is similar to the surface free energy of the substrate and, therefore, can be neglected. Calculating the reduction of the exposed Bi surface by:

$$S_{cylinder} - S_{cap} \quad (8)$$

we find a reduction of approximately 22.000 nm², equal to a total energy-gain of 55 keV or 4.2 meV/atom using a Bi density of 28 atoms/nm² [79]. Although dewetting the film into spherical caps results in an energy gain of 4.2 meV/atom, the activation energy seems to be reached upon the melting temperature (550 K approximately 4.7 meV/atom).

When performing the same grain analysis for increasing film thickness grown by PLD, we can easily verify whether our assumption for the cylinder-shaped grains in the as-grown films is valid. In Fig. 12(a) and (b) we plotted the obtained <a/2> and <h> data for the melted and solidified spherical caps. In Fig. 12(b), the film thickness appears to have a linear relation to the resulting height of the spherical cap, whereas its diameter appears to depend as $\sqrt{\langle a/2 \rangle}$ on the grown film height. The latter also represents the interface between Bi and substrate ($\pi$) to linearly depend on the film height grown. As this relation is linear with film height as well as <h>, it confirms our assumption of conservation of the shape and aspect ratio.

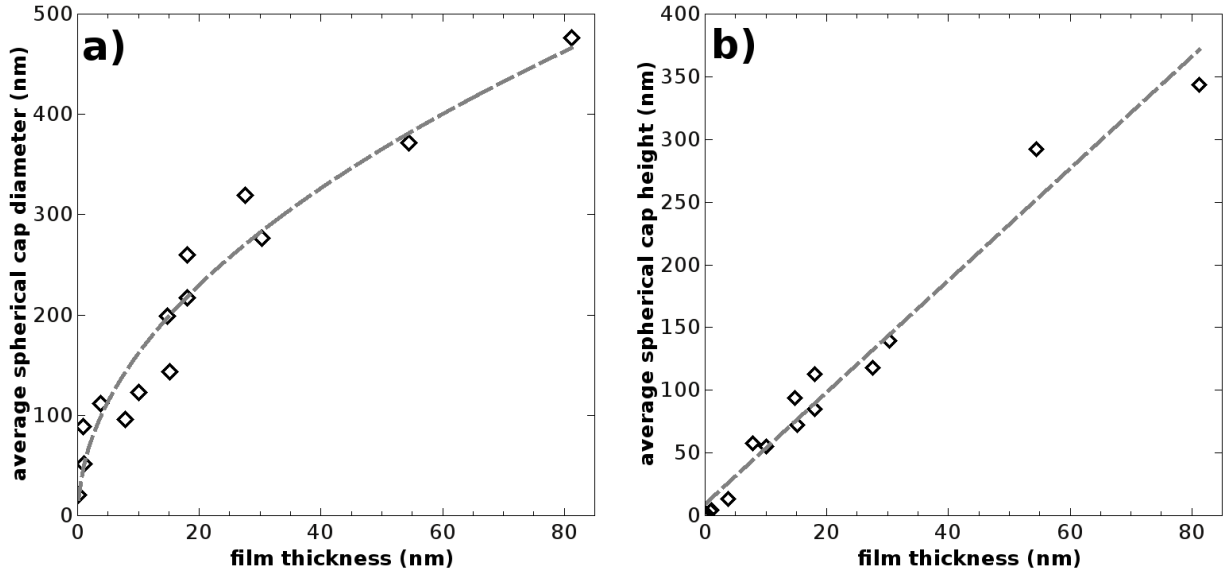

*Figure 12: (color online) (a) The average spherical cap diameter as a function of the deposited film thickness. (b) The average spherical cap-height h as a function of the deposited film thickness.*

This brings us to the growth, melting, and solidification model shown in Fig. 13(a). As we previously described [39], at low temperature, the deposition of Bi results in (thick) rougher Bi(110) films, illustrated in cartoon Fig. 13(a)(i). Upon annealing towards 500 K, these films show a phase transition towards a Bi(111) film, Fig. 13(a)(iii), where the surface roughness is decreased to sub-Å, Fig. 13(a)(iv). When the film is melting, Wulff shaped crystals appear as a result of dewetting, Fig. 13(a)(v), ultimately dewetting the substrate, Fig. 13(a)(vi) in agreement with High-Resolution Transmission Electron Microscopy (HRTEM) observations in literature where a freezing mechanism is mediated by crystallization of an intermediate ordered liquid [46], [47], [80]. Cooling a melted film results in solidification as the critical temperature is hit upon, the first nucleation at the surface starts, followed by ultrafast lateral growth up to fixed sizes. Once the lateral dimensions of the grain are set, further crystallization proceeds in height resulting in Bi(110) spherical caps, Fig. 13(a)(vii). These observations are supported by recent observations of the crystallization of a Bi nanodroplet shown in Fig. 3 of Ref. [80].

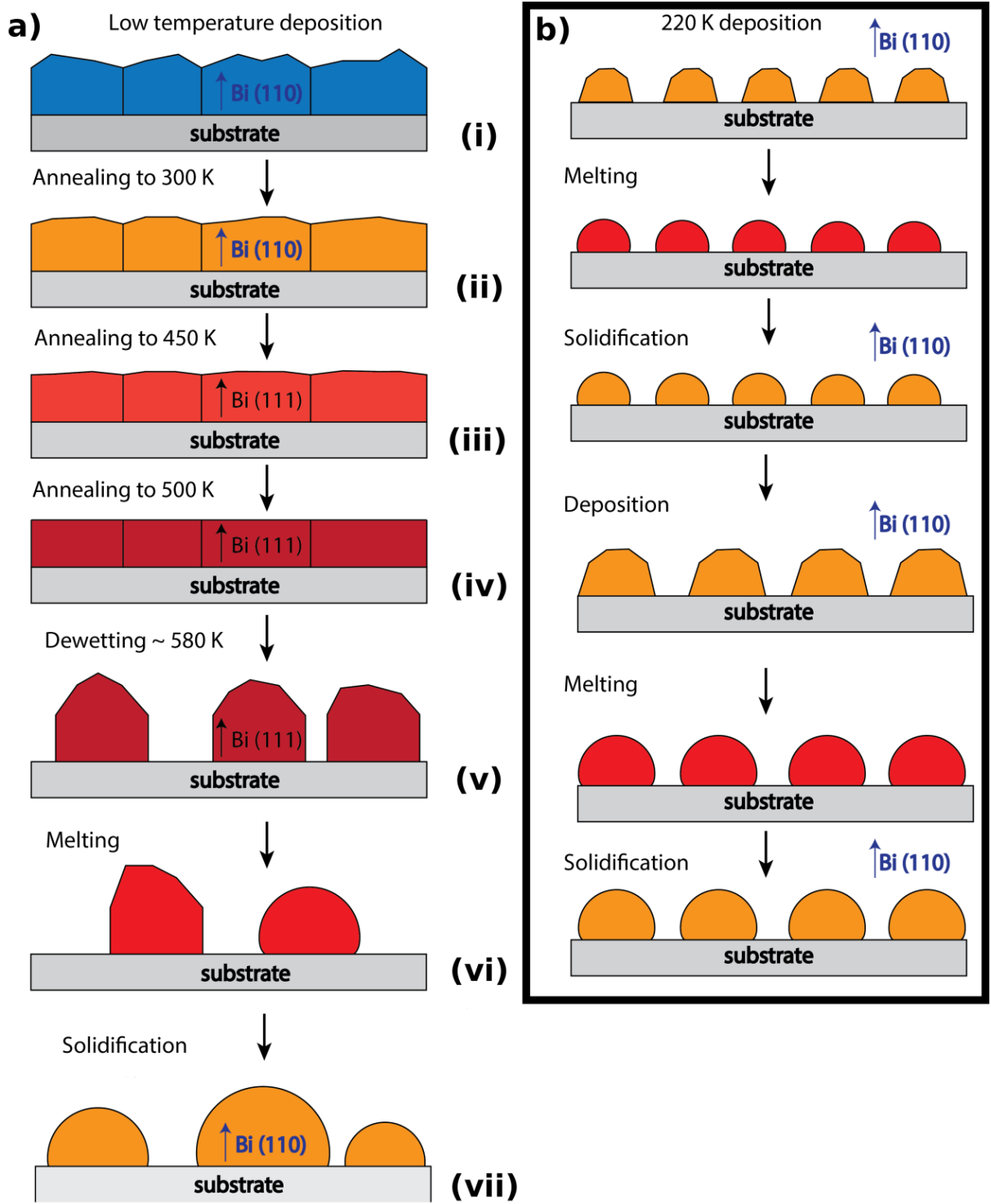

*Figure 13: (color online) (a) Cartoon impression of the resulting Bi(110) structures upon growth at LT (i), annealing (ii), phase transition towards Bi(111) (iii), flattening (iv), dewetting (v) and melting. Upon cooling the melted film, solidification results in Bi(110) spherical caps (vi). (b)*

*Multiple sequences of deposition at elevated temperature followed by melting can be used as a route towards enlarging and controlling the spherical cap size. The arrows mark the perpendicular direction to the labeled crystallographic plane parallel to the substrate.*

As we previously described [39], growth at higher temperatures, such as, e.g., 220 K, results in rougher Bi(110) films; see also the cartoon impression in Fig. 13(b)(i). Deposition of Bi on these films, followed by melting and solidification (see Supplementary Movie M3 where deposition of the melting and solidification of nanodots is presented), results in well-defined nanodots, the spherical caps, see Fig. 13(b), which separation can be derived from in-plane profiles of the Bi(110) Bragg presented in Fig. 14 (a). The profiles show the sharp and intense central Bragg peak and two lobs at its side, well visible after subtraction of the central peak, see Fig. 14 (b) [81]. These lobs originate from diffuse scattering from the X-ray waves scattered from evenly separated Bi nanodots present at the surface. The position of the lob centers corresponds to the average separation distance between the Bi nanodots. The evolution of the separation distance as a function of increasing film thickness is presented in Fig. 14 (c). The initial separation of the Bi nanodots follows a linear trend up to a (nominal) coverage of ~2 nm above which we observe a steep increase until the coverage reaches 5 nm. This can be interpreted as an initial steady growth of small spherical caps, which merge into larger entities above coverage of ~2 nm by the Ostwald ripening or collective diffusion as whole clusters migrate over the surface. Above a coverage of 5 nm, the merging process slows down for an average separation distance above 40 nm. Further deposition, growth, melting, and solidification of the film increase the spherical capsize as the already grown nanodots act as nucleation sites for the newly deposited atoms. Therefore, this scheme can be a viable route towards the size-controlled growth of Bi nanodots with a well-defined separation distance. As we have shown, by selecting the appropriate initial film thickness and subsequently heating and cooling this film, one can tune the size and separation of the Bi spherical caps.

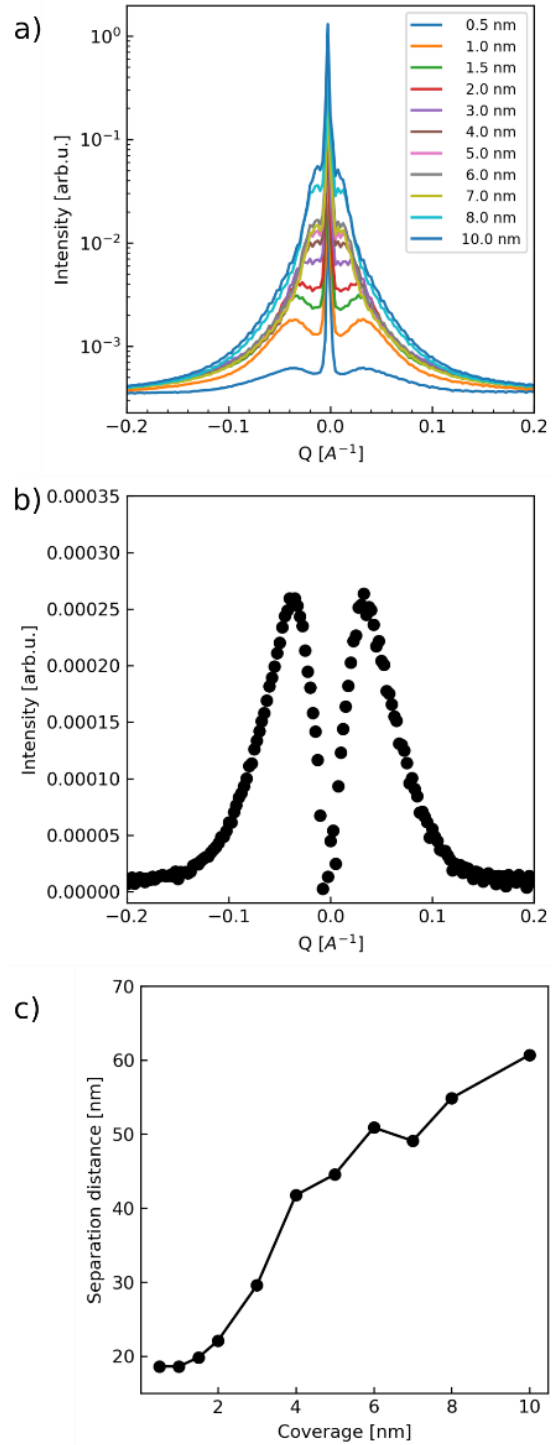

*Figure 14: a) The Bi 110 Bragg peak profiles composed of the central Bragg peak part and diffuse scattering visible as side lobs on both sides of the peak, recorded as a function of the film thickness.*

*b) The Bi 110 Bragg peak profile recorded at 0.5 nm coverage after subtracting the central peak. c) The average separation distance between the Bi spherical caps on the surface derived from the position of the side lobs presented in a).*

As discussed in the literature, the bandgap of the Bi particles on such an insulating substrate can be tuned by about 2.5 eV [69]. To verify the feasibility of this approach, we measured by amplitude modulated (AM)-KPFM the local contact potential difference (LCPD) of such spherical caps, as shown in Fig. 15(b). Plotting the LCPD versus the inverse cap diameter, we find a linear relation as plotted in Fig. 15(c) for the data obtained from Fig. 15(b). This confirms the observed workfunction relation for nanodots to be inversely linearly proportional to the number of electrons confined [82].

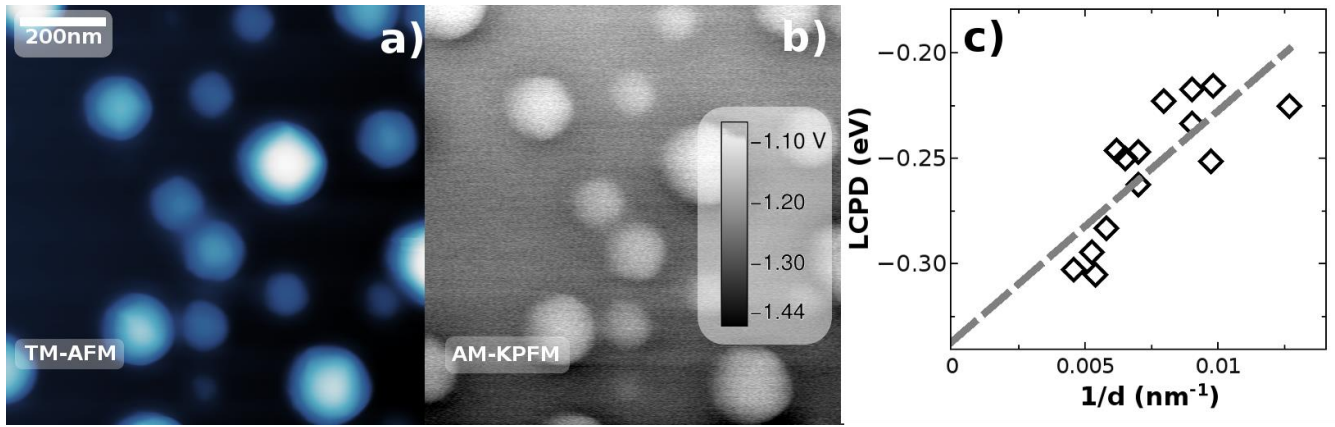

*Figure 15: (color online) (a) TM-AFM and (b) AM-KPFM image of Bi spherical caps on the sapphire substrate, revealing slight variations in the LCPD. (c) The LCPD plotted against the inverse cap diameter reveals a linear relation.*

## Conclusions

We have described multiple routes toward the controlled growth of Bi nanostructures using PLD and MBE techniques. The size of Bi nanodots can be controlled by the initial film thickness of the deposited bismuth film, followed by subsequent melting and solidifying. The nanodots take the shape of cylinders and spherical caps, which is attributed to the energetical balance in dewetting between the substrate and thin-film material. The in situ and real-time monitoring of the Bi melting allows for a real-time investigation of the crystallographic orientation change and sheds light on Bi nanostructures' melting and solidification kinetics. Deposition at low temperatures ~40 K leads to the formation of uniform Bi (110) films, and subsequent rise of the temperature leads to their smoothening and finally change of the crystal structure towards Bi(111). Just before reaching the melting point, Bi films show rapid dewetting into nanodots and subsequently liquefy. The solidification of the nanodots is observed well below the melting point of Bi due to its supercooled state. The liquid Bi solidifies into nanodots oriented with the (110) crystallographic plane perpendicular to the substrate. The direct deposition above 220 K leads to the formation of faceted nanocrystals having the (110) orientation. The presented results show a plethora of different scenarios for the growth of controlled Bi nanostructures, thereby controlling the shape, size, and crystallographic orientation.

MJ and TRJB would like to thank Helena Isern and Thomas Dufrane for their technical assistance. This work is part of the Foundation for Fundamental Research on Matter (FOM) research program, which is part of the Netherlands Organisation for Scientific Research (NWO).  TRJB is grateful to Prof. Dr. Ing. G. Rijnders for his support on the experimentally obtained data.

**Supplemental Materials: Bismuth growth kinetics during liquification and solidification.**

The Supplemental Material contains Tapping Mode Atomic Force Microscopy (TM-AFM) images, optical microscopy images, X-ray reflectivity, and diffraction (XRD) data, Scanning Electron Microscopy (SEM) images, and plots of the Scherrer constant *K* dependence.

In Fig. 15 we plotted the X-ray reflectivity and diffraction curves for the Bi films grown at decreasing ablation energy density of 5, 4, 3, 2, and 0.7 J/cm$^2$ for 500 subsequent pulses. Modeling and fitting the reflectivity curves in Fig. 15(a), we find film thicknesses of respectively 14.7 nm, 12.4 nm, 10.1 nm, 7.9 nm, and 7.4 nm. All films reveal a roughness below 1 nm, as also discussed below. As discussed in the literature [49], increased ionic energy would result in changing the Bi crystal structure of Bi(111) at low ionic energy (~110 eV) towards Bi(110) crystal structures for ionic energies of about 270 eV. Although, as in our setup, we make use of an excimer laser with a 25 ns pulse duration, in contrast to the Nd:YAG laser having a pulse duration of 5 ns in Ref. [49], and we do not select the ion energy by the use of a Langmuir planar probe [49], we do not find any change in the crystal structure of our grown films as shown in Fig. 15(b).

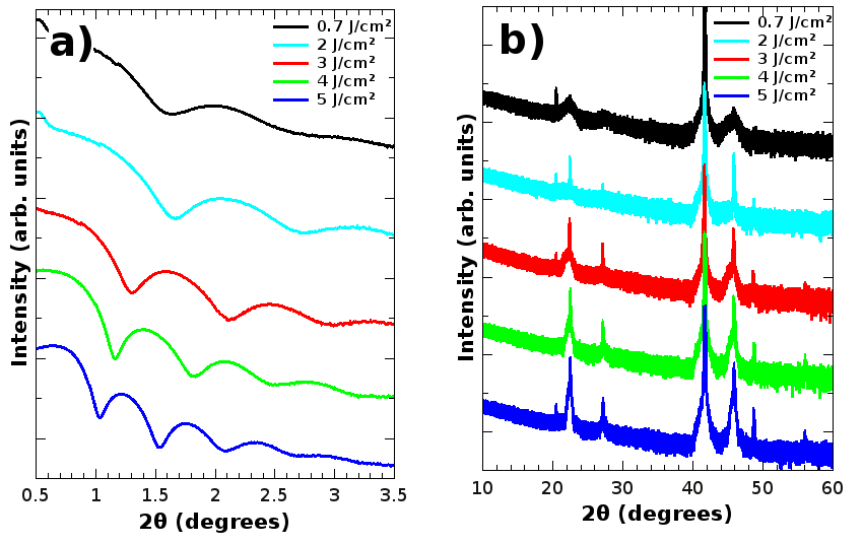

*Figure 15: a) X-ray reflectivity curves for Bi films grown at varying ablation energy density for 500 subsequent pulses. b) the corresponding X-ray diffraction data reveal a similar ratio mixture of Bi(111) and Bi(110) crystal structures, independent of ablation energy density used.*

To select an appropriate ablation energy density, we measured the morphology for all films grown at varying ablation energy densities, as shown in Fig. 16(a-e). Measuring their roughness (RMS), we find values of 0.90 nm, 0.76 nm, 0.63 nm, 0.73 nm, and 0.58 nm for an ablation energy density of respectively 5, 4, 3, 2, and 0.7 J/cm$^2$. From this, one would be tempted to conclude that the ablation energy density seems rather irrelevant for the deposited film quality. However, compared to the PLD of oxides, the ablation of metal targets requires ablation thresholds typically an order of magnitude higher [65]. However, too large ablation thresholds in combination with a too low process pressure result in droplet formation at growth [60-64]. Furthermore, a study on the microstructural evolution of Ni films upon increasing ablation threshold [83] revealed an increased clustering and denser packing of nanoparticles. Besides this, it should be noted that to reduce the droplet density, a (material dependent) sweet spot exists, as discussed in Fig. 5 of Ref. [63]. In Fig. 16(f-j), we show optical microscopy images corresponding to the TM-AFM morphologies of Fig. 16(a-e). From these images, the increased droplet formation for increased ablation energy density can be easily seen. Now, to have a film with low surface roughness, well-defined crystal structure, and minimal droplet formation, we decided to stay within the energy range of 0.7-2 J/cm$^2$.

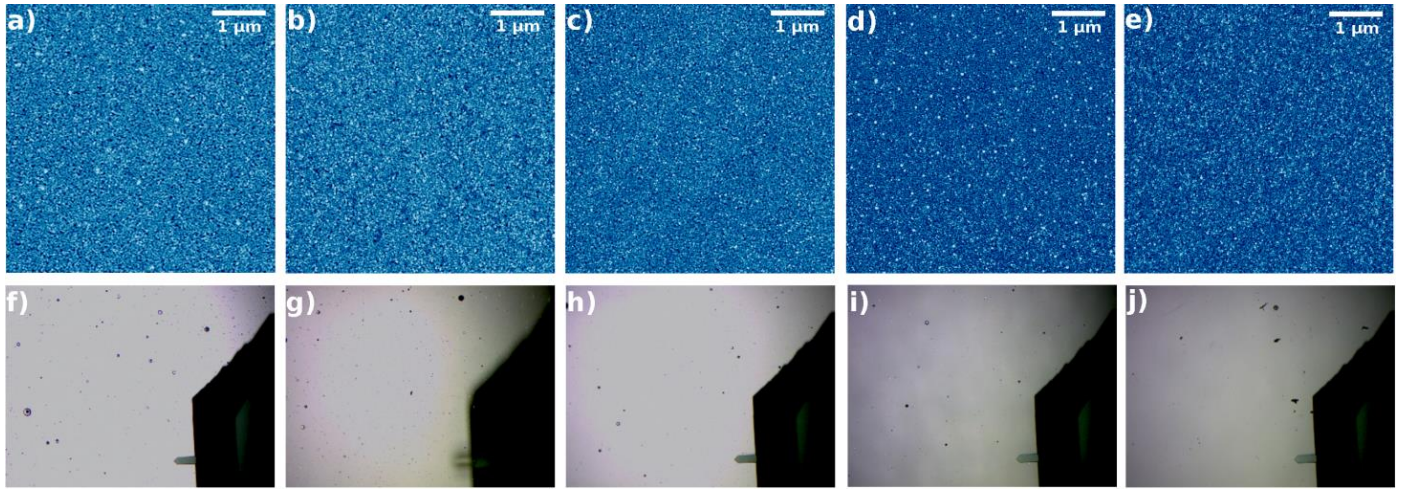

*Figure 16: (color online) TM-AFM images for the Bi films grown at a decreasing ablation energy density of 5 J/m$^2$ (a), 4 J/m$^2$ (b), 3 J/m$^2$ (c), 2 J/m$^2$ (d), 0.7 J/m$^2$ (e). The corresponding optical microscopy images revealing droplet formation typical for PLD growth of metal films are shown below (f-j).*

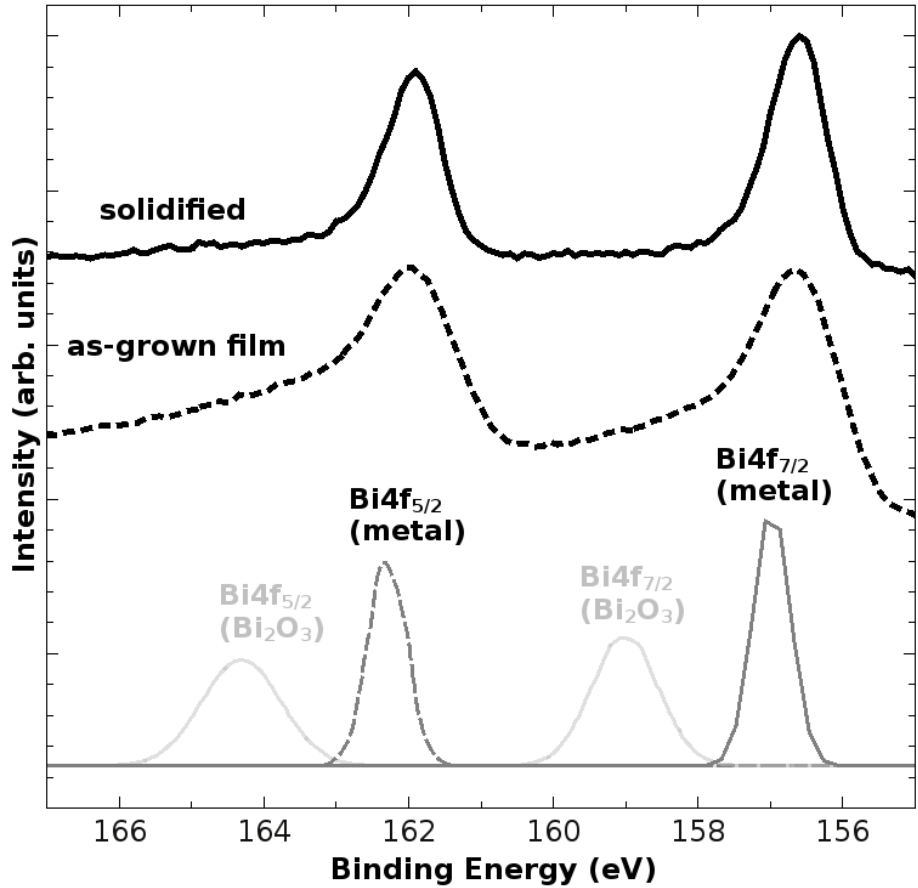

*Figure 17: (a) XPS curve of the (PLD) as-grown Bi film and the solidified spherical caps. Inspection of the Bi $4f_{5/2}$ and Bi $4f_{7/2}$ peak position reveals the metallic state of the film and the spherical caps.*

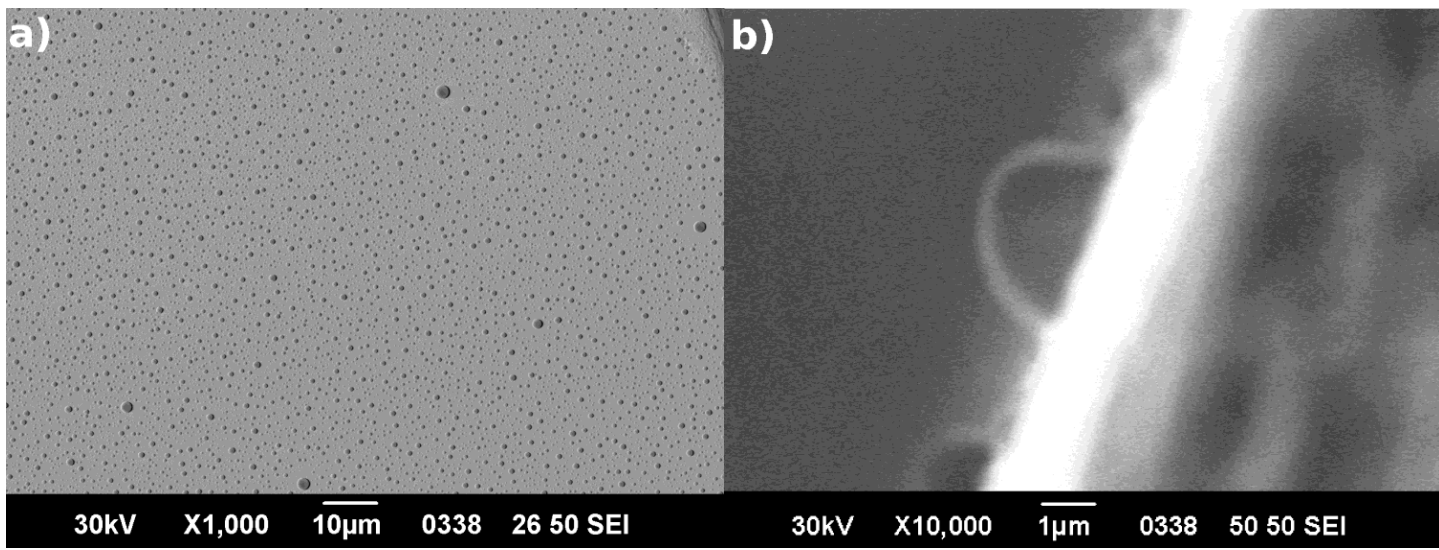

*Figure 18: (a) SEM image of a cooled 81.8 nm thick Bi film, corresponding to Fig. 3(a). (b) SEM image (side view) of a larger Bi spherical cap.*

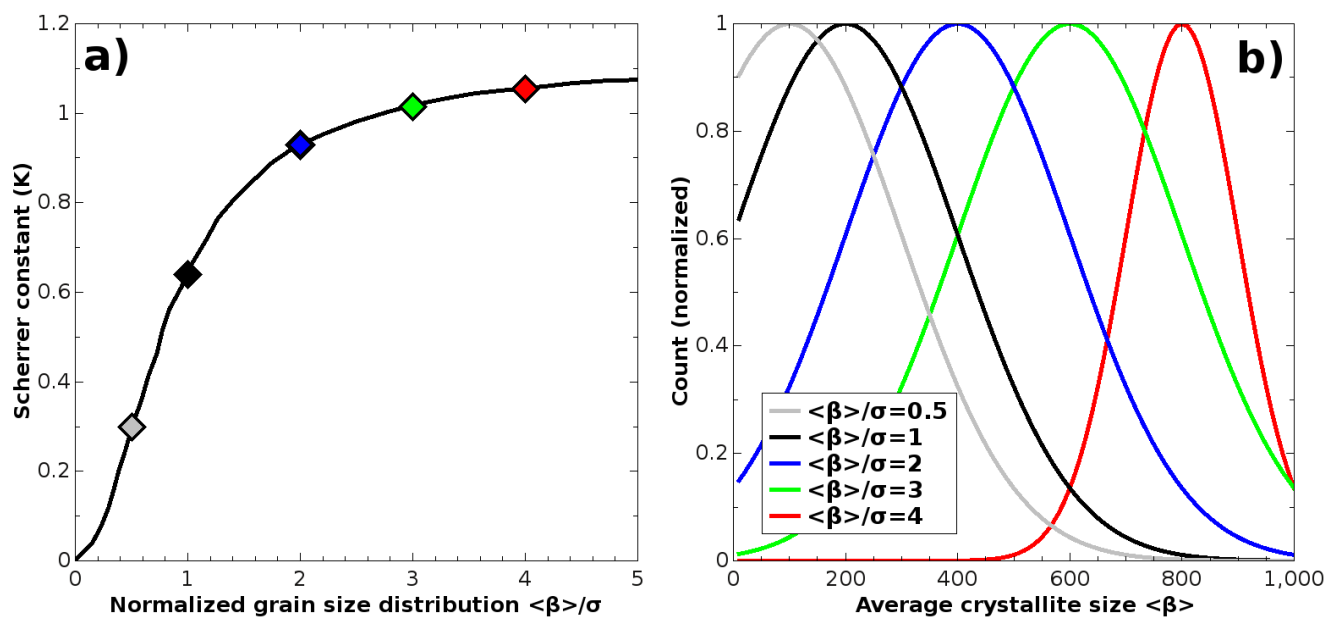

*Figure 19: (color online) (a) The dependence of the Scherrer constant on the normalized grain size distribution ($\beta/\sigma$). The diamond markers correspond to the normalized size distributions in (b), where the average crystallite size ($\beta$) increases for constant grain size distribution width ($\sigma$), resulting in $\beta/\sigma$=0.5, 1, 2 and 3. The curve for $\beta/\sigma$=4 has a narrowed grain size distribution width.*